\documentclass[11pt, a4paper]{article}
\usepackage{moriond,epsfig}
\def\be{\begin{equation}}
\def\ee{\end{equation}}
\def\bea{\begin{eqnarray}}
\def\eea{\end{eqnarray}}

\begin{document}
\vspace*{4cm}
\title{FORWARD-BACKWARD CHARGE ASYMMETRY IN Z PRODUCTION AT THE LHC.}

% Changed 040623 - All authors go here
\author{ Mohamed Aharrouche}
\address{Laboratoire d'Annecy-le-Vieux de Physique des Particules\\
Universit\'e de Savoie, CNRS/IN2P3, Annecy-le-Vieux CEDEX France}

\maketitle \abstracts{ We present here a study on the
determination of the effective weak mixing angle,
$\sin^{2}\theta^{lept}_{eff}$ from the measurement of the
Forward-Backward Asymmetry with a high a statistical precision,
10$^{-4}$. To reach such a precision it is necessary to identify
the electrons in the forward regions of the ATLAS detector. It is
demonstrated that one can reach an electron-jet rejection of more
than 100 with an efficiency on electron reconstruction better than
50$\%$, by using a multivariate analysis.}

\noindent {\small¥{\it Keywords}: LHC, ATLAS, Z$\rightarrow$e+e-,
Forward-Backward Asymmetry, $\sin^{2}\theta^{^{lept}}_{eff}$.}

\section{Introduction}

Many measurements of the theoretical electroweak parameters are
sensitive to the quantum corrections effect associated to the
Higgs boson. The W mass, $M_{W}$, and the weak mixing angle,
$\sin^{2}\theta^{^{lept}}_{eff}$, are related quadratically to the
top mass and logarithmically to the Higgs mass. Therefore the
indirect measurement of the Higgs mass from the top mass and the
weak mixing angle provides a check of the Standard Model coherence
and a validation of any Higgs discovery.

At the LHC, a Z production rate of $\sim$1.5$\times$10$^8$ per
year at high luminosity of which $\sim$5$\times$10$^6$ decay to an
electron-positron pair is expected. The determination of the weak
mixing angle from the measurement of the forward-backward
asymmetry in the $Z\rightarrow e^+e^-$ with a very small
statistical error comparable to the LEP one is possible. The
electron channel was chosen instead of the muon due to the limited
coverage of the muons (up to $|\eta|$= 2.7).

\section{Simulation}

The signal and background events are generated with the PYTHIA6.2
and the parametrization CTEQ5L of the structure function is used.
The events are then fast simulated and reconstructed with the
ATLFAST, a fast simulation of the ATLAS detector response.

The events $q\overline{q}\rightarrow Z/\gamma^*\rightarrow e^+e^-$
were generated in various transverse momentum ($\hat{p}_T$) ranges
of the hard scattering matrix element and with a dilepton mass,
$\hat{m}$, greater than 50 GeV.

The main backgrounds to the electron channel are:

\begin{itemize}
\item $pp\rightarrow jj$ (QCD): it is the dominent background
where each jet simulates an electron. The cross section of this
processus is greater by several orders of magnitude than the
signal one, and dominates at low transverse momentum.

\item $pp\rightarrow t\overline{t}\rightarrow e^+e^-$: The top
quark decay into the $W$ boson and b quark, followed by the $W$
decay into electron and neutrino ($t\rightarrow Wb$, $W\rightarrow
e\nu$). It has the same signature as the signal one as the two
electrons of the final state can simulate the two electrons from
Z.
\end{itemize}

\section{Analysis method}

The aim of this analysis is the measurement of the
forward-backward charge asymmetry in the $Z\rightarrow e^+e^-$
events and its precision. This measurement provides a
determination of the weak mixing angle
$\sin^{2}\theta^{lept}_{eff}$ by using the
relation\cite{jon,baur}:
\begin{eqnarray}
A_{FB}=b(a-\sin^{2}\theta^{lept}_{eff})
\end{eqnarray}\label{sin}

The selection cuts of our analysis in the electron channel
requires an electron transverse momentum higher than 20 GeV
($p_T>$ 20GeV) to simulate the energy threshold of the electron
trigger, and a window of 12 GeV around the Z mass, 85.2 GeV $<$
M$_{(e^+e^-)}$ $<$ 97.2 GeV (Z pole). One requires that one of the
two electrons lies in the central region ($|\eta|<$ 2.5), while
the other electron is either in the central region (case 1) or in
the forward region (case 2) up to $|\eta|$ = 4.9. In the region
2.5$<|\eta|<$ 3.2 the calorimeters used are the EMEC and the HEC
and for $|\eta|>$ 3.2 the forward calorimeter (FCal) is used. Note
that we can't reconstruct the electron track in the forward region
(2.5 $<|\eta|<$ 4.9) as the tracking system of ATLAS is limited to
the region $|\eta|<$ 2.5. In addition, the forward calorimeters
have a coarser granularity than in the central one and we expect
the electron identification in this region to be less performant
than in the central one.

At low rapidity y$_{(e^+e^-)}$ most of the events are produced via
the annihilation of the sea quark and sea anti-quark, and the
probability that the valence quark and the di-electron boost
coincide is then lower. The effect on a cut $|y_{(e^+e^-)}|>$1 is
then studied.

We require the missing transverse energy to be less than 20 GeV
($P_T^{miss} <$ 20 GeV). This cut rejects efficiently the
background coming from $pp\rightarrow t\overline{t}$ channel where
the top decays semileptonically.

In practice the forward-backward asymmetry is calculated by a
counting method of the forward events $N_F$ with $\cos\theta^*>$ 0
and backward events with $\cos\theta^*<$ 0 ($\theta^*$ is the
polar angle of the electron in the Z rest frame):
\begin{equation}
A_{FB} = \frac{N_F-N_B}{N_F+N_B}
\end{equation}
As the distribution of the events $N_F$ and $N_B$ follows a
binomial distribution, the error on the two quantities can be
written as follows:
$\sigma_{N_F}=\sigma_{N_B}=\sqrt{N_FN_B}/\sqrt{N_F+N_B}$ and the
$A_{FB}$ error is $\sigma_{A_{FB}} = \sqrt{\frac{1-A_{FB}^2}{N}}$.

In this analysis, the electron/jet rejection is studied for a
fixed electron efficiency (50$\%$). The data are normalized to the
integrated luminosity of 100 fb$^{-1}$ (3 years at high
luminosity).

\section{Results}

\begin{figure}[h]
%\begin{center}
\includegraphics[width=8.5cm,height=6cm]{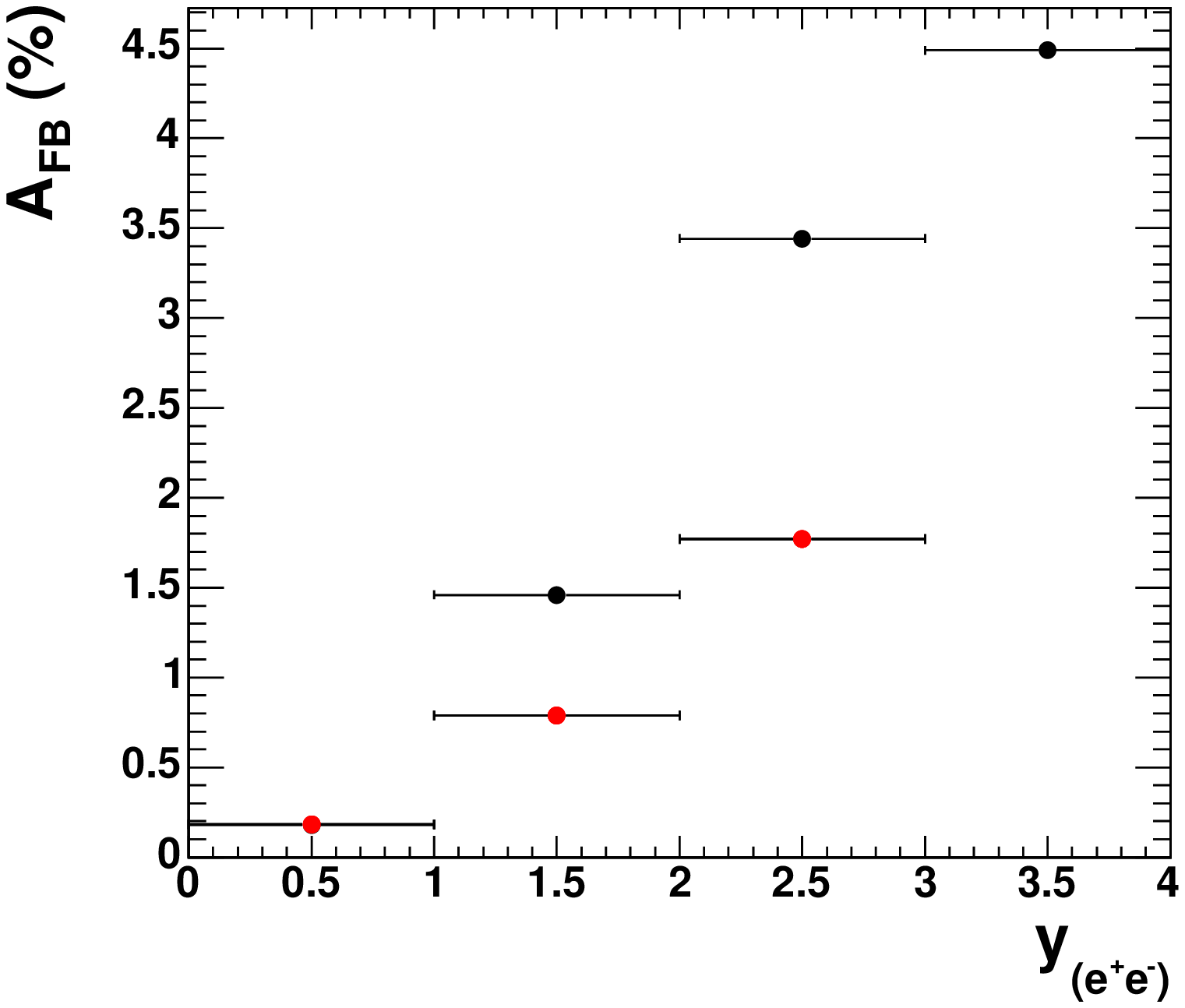}\hfill
\includegraphics[width=8.5cm,height=6cm]{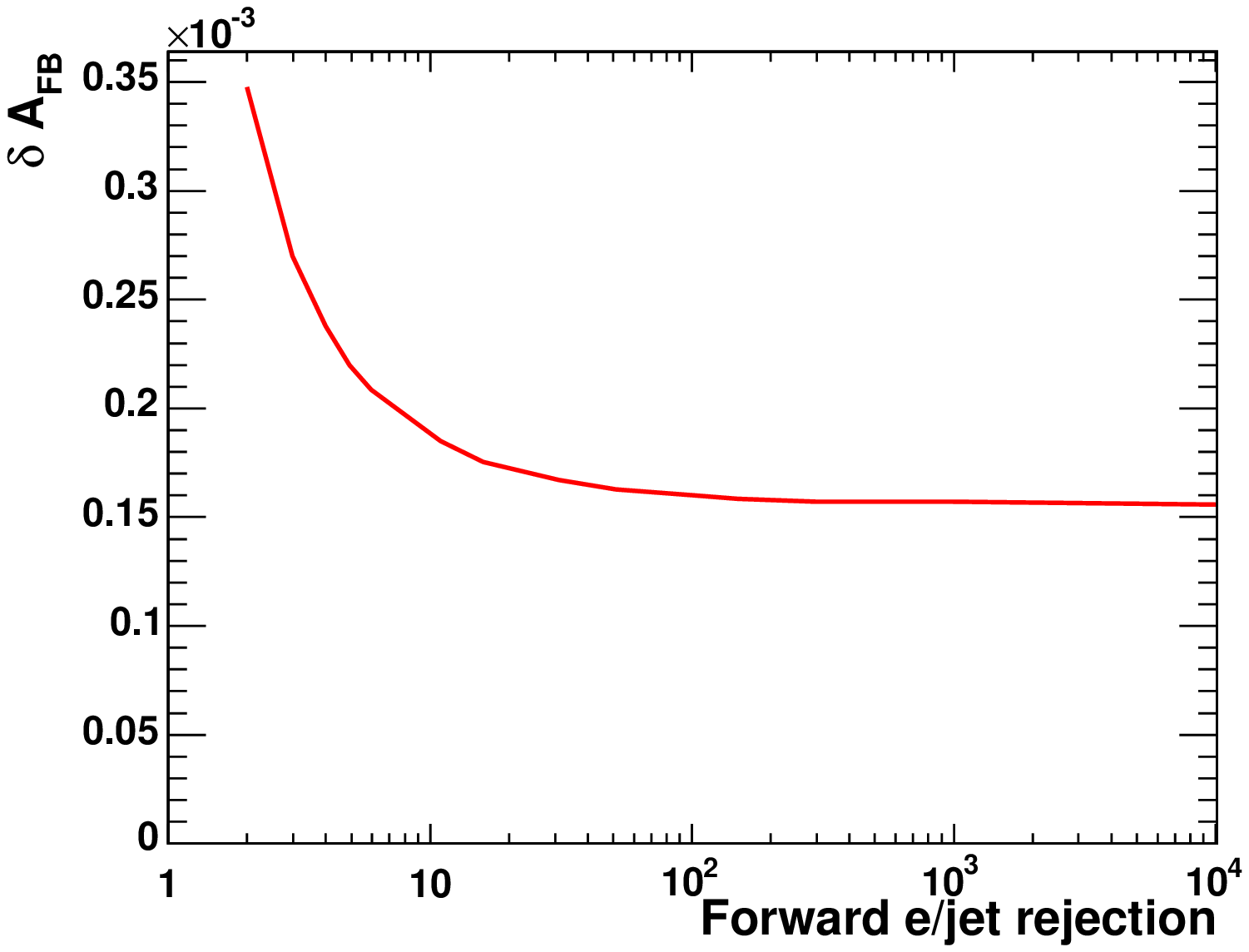}
\caption{\it Left: Forward-Backward asymmetry versus dilepton
rapidity in the case 1 (red points) and in the case 2 (black
points). Right: Forward-Backward asymmetry accuracy versus the
forward electron/jet rejection in the events of the case 2.}
\label{afb}
%\end{center}
\end{figure}

Table \ref{dat2} shows the value of the forward-backward asymmetry
measurement, its statistical error and the corresponding error on
the weak mixing angle $\sin^{2}\theta^{lept}_{eff}$. When the two
electrons are in the central region, we remark that the asymmetry
and its error are unchanged with or without the background due to
the higher rejection factor in this region. Fig.\ref{afb} left
shows the variation of asymmetry versus the rapidity of the two
electrons. It is observed that the asymmetry increases by a factor
2 when allowing the second electron to be in $|\eta|<$4.9. As
shown in the right plot of Fig.\ref{afb}, the accuracy on the
forward backward asymmetry improves while the jet rejection
increases in the forward regions and it is almost constant for
rejection greater than 100. The statistical error reached here
(for a forward rejection of 100 and forward electron efficiency of
50$\%$) on the weak mixing angle is of $\sim$10$^{-4}$.

The statistical error on $\sin^{2}\theta^{lept}_{eff}$ is deduced
from the relation \ref{sin} where the parameters, a and b, are
derived from theory\cite{baur} including the radiative
corrections.

\begin{table}[h]
\begin{center}
    \begin{tabular}{lllll}
            \hline
               & $Rej_{fwd}$ & A$_{FB}$ ($\%$)&$\delta$A$_{FB}$ & $\delta \sin^{2}\theta^{lept}_{eff}$ \\
            \hline
              $|y_{e^-}|$,$|y_{e^+}|<$ 2.5 &signal &0.59 &  1.35 $\times 10^{-4}$&4.35 $\times 10^{-4}$\\
              &sig+bkgd&0.59&  1.35 $\times 10^{-4}$&\\
              \hline
              $|y_{e^-}|$,$|y_{e^+}|<$ 2.5,&signal&1.13 & 1.96$\times 10^{-4}$&2.58 $\times 10^{-4}$ \\
               $|y_Z|>$1&&&\\
               &sig+bkgd&1.13& 1.96$\times 10^{-4}$&\\
               \hline
             $|y_{e_1}|<$2.5, &signal& 1.29 & 1.19$\times 10^{-4}$ & 0.97 $\times 10^{-4}$\\
             $|y_{e_2}|<$ 4.9&&&&\\
             &10$^4$&1.29 & 1.19$\times 10^{-4}$&0.97 $\times 10^{-4}$\\
             &100&1.26 & 1.21$\times 10^{-4}$&0.98 $\times 10^{-4}$\\
             &10&1.05 & 1.33$\times 10^{-4}$&1.08 $\times 10^{-4}$\\
             &1&0.39 & 2.2 $\times 10^{-4}$&1.79 $\times 10^{-4}$\\
             \hline
             $|y_{e_1}|<$2.5,&signal&  2.12 & 1.56 $\times 10^{-4}$&0.96$\times 10^{-4}$\\
             $|y_{e_2}|<$ 4.9,$|y_Z|>$1 &&&&\\
             &10$^4$&2.12 & 1.56 $\times 10^{-4}$ &0.96$\times 10^{-4}$\\
             &100&2.03 & 1.59 $\times 10^{-4}$ &0.97$\times 10^{-4}$\\
             &10&1.52&1.84 $\times 10^{-4}$ & 1.13$\times 10^{-4}$\\
             &1&0.43 & 3.47 $\times 10^{-4}$ &2.13$\times 10^{-4}$ \\
            \hline
    \end{tabular}
\caption{\it Values of the FB asymmetry, of the statistical error
on the asymmetry and on the weak mixing angle for 4 rejection
values in the forward region (10$^4$, 100, 10 and 1).}
\label{dat2}
\end{center}
\end{table}

\section{Forward electron reconstruction}

\begin{figure}[h]
%\begin{center}
\includegraphics[width=6.5cm,height=5cm]{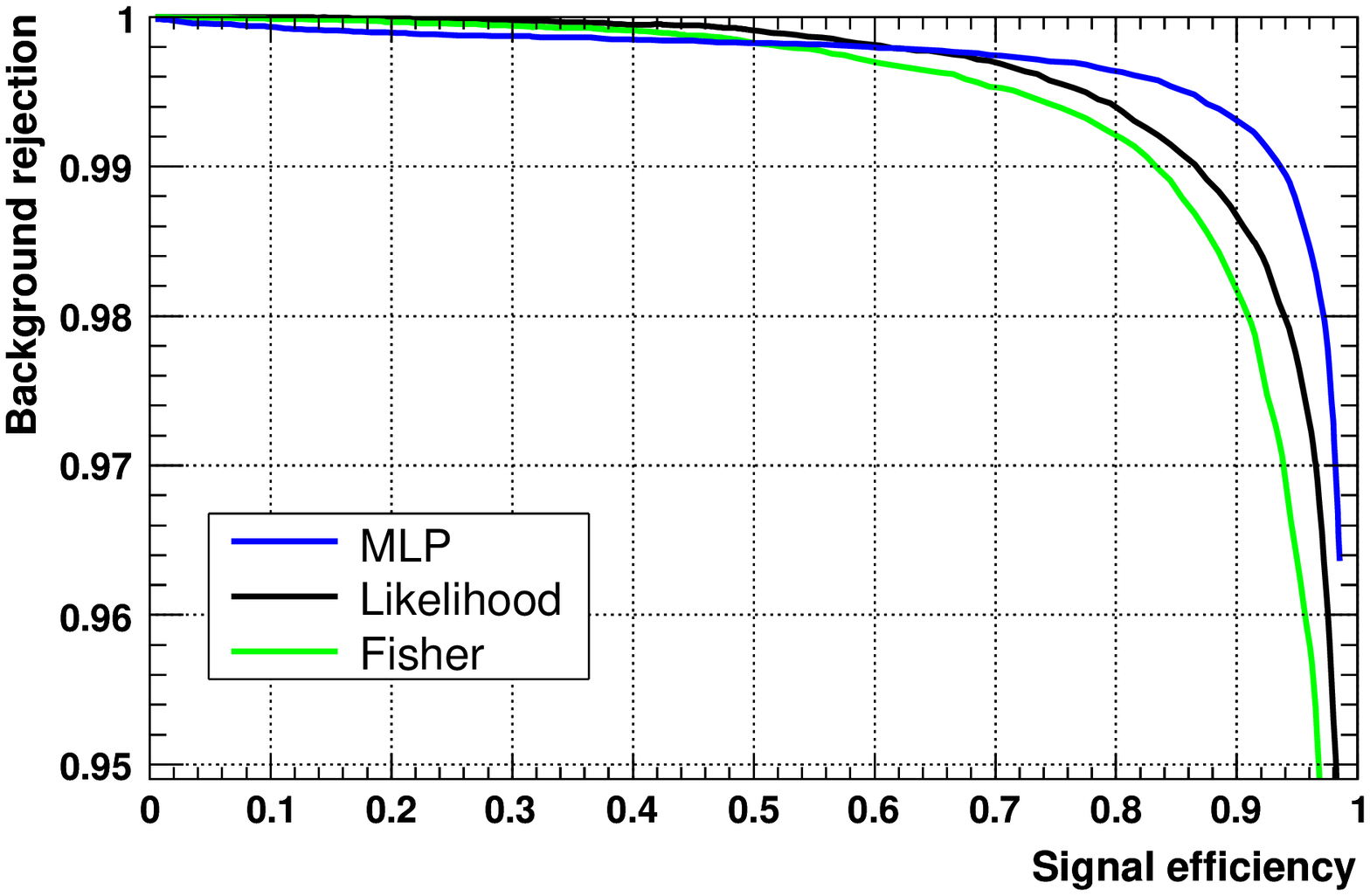}\hfill
\includegraphics[width=6.5cm,height=5cm]{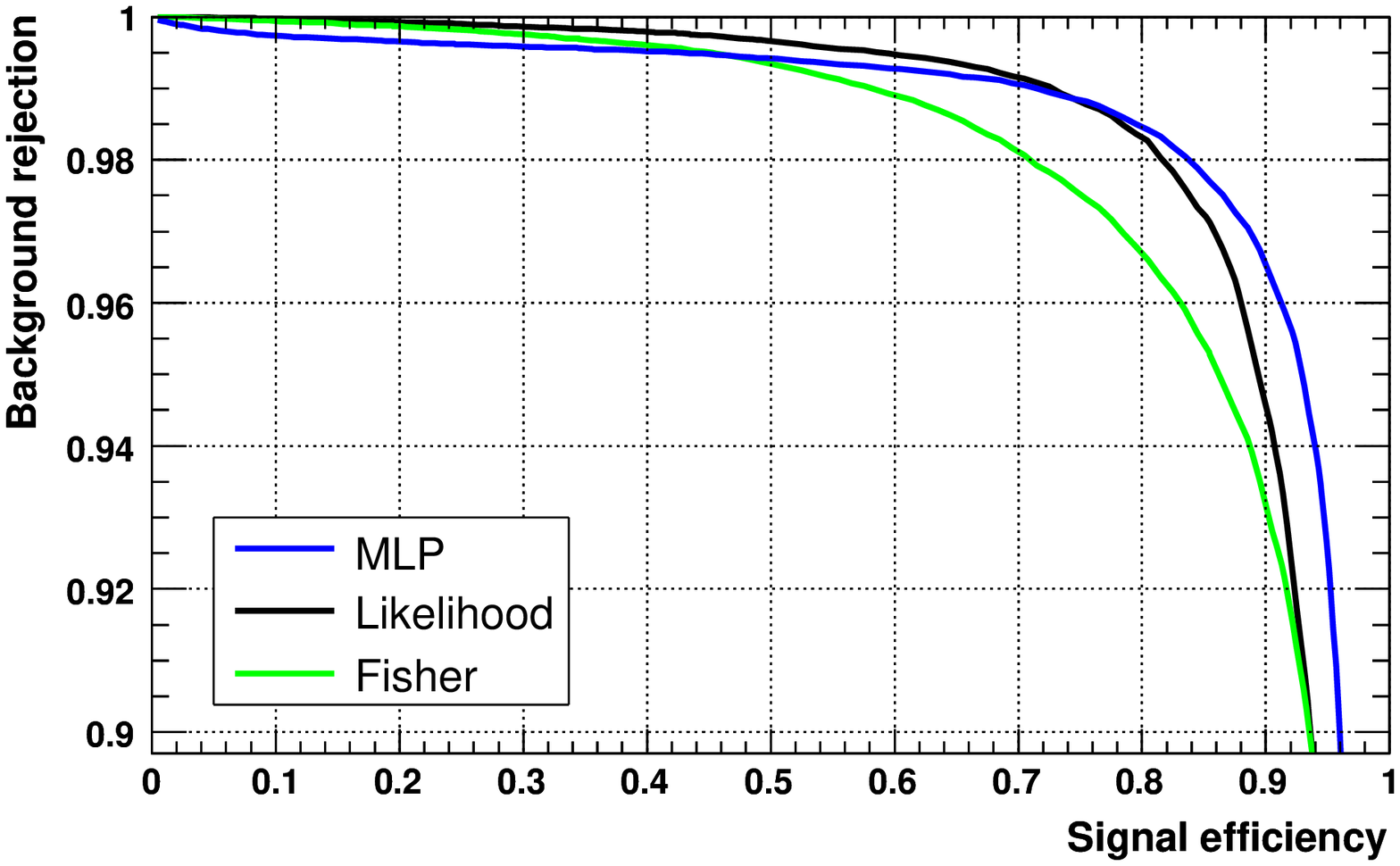}
\caption{\it Background rejection versus the signal efficiency in
the EMEC (left) and in the forward calorimeter (right).}
\label{eff}
%\end{center}
\end{figure}

In order to evaluate our analysis result and demonstrate if we can
reach such requirements defined by our analysis, a multivariate
analysis was used to valuate the performance of the forward
calorimeters in the separation of electrons from hadrons. The
input variables used describe the shower shape (lateral and
longitudinal) and its development. The most discriminating
variable is the fraction of the total energy deposited in the most
energetic cell. Various discriminant methods are used to confirm
the obtained result. Fig. \ref{eff} shows the variation of the jet
rejection versus electron efficiency in the forward regions for
three different analysis methods (Fisher discriminant, maximum
likelihood, neural net). We found that a rejection of 100 can
easily be obtained while keeping an electron efficiency better
than the 50$\%$.

\end{document}